\documentclass[sigconf]{acmart}

\usepackage{booktabs} 
\usepackage{graphicx, amsmath, amssymb, amsfonts, mathrsfs}

\setcopyright{rightsretained}

\acmDOI{10.475/123_4}

\acmISBN{123-4567-24-567/08/06}

\acmConference[KDD'18]{ACM KDD conference}{Fragile Earth Workshop}{.}
\acmYear{2018}
\copyrightyear{2018}

\acmArticle{4}
\acmPrice{15.00}

\begin{document}
\title{Agriculture Commodity Arrival Prediction using Remote Sensing Data: Insights and Beyond}


\author{Gautam Prasad}
\affiliation{%
  \institution{Microsoft India Dev. Centre (IDC)}
  \streetaddress{P.O. Box 1212}
  \city{Hyderabad} 
  \country{India}
}
\email{gapras@microsoft.com}

\author{Upendra Reddy Vuyyuru}
\affiliation{%
  \institution{Microsoft India Dev. Centre (IDC)}
  \city{Hyderabad} 
  \country{India} 
}
\email{upendrav@microsoft.com}

\author{Mithun Das Gupta}
\affiliation{%
  \institution{Microsoft India Dev. Centre (IDC)}
  \city{Hyderabad} 
  \country{India} 
}
\email{migupta@microsoft.com}

\renewcommand{\shortauthors}{G. Prasad et al.}

\begin{abstract}
In developing countries like India agriculture plays an extremely important role in the lives of the population. In India, around 80\% of the population depend on agriculture or its by-products as the primary means for employment. Given large population dependency on agriculture, it becomes extremely important for the government to estimate market factors in advance and prepare for any deviation from those estimates. Commodity arrivals to market is an extremely important factor which is captured at district level throughout the country. Historical data and short-term prediction of important variables such as arrivals, prices, crop quality etc. for commodities are used by the government to take proactive steps and decide various policy measures.

In this paper, we present a framework to work with short timeseries in conjunction with 
remote sensing data to predict future commodity arrivals. 
We deal with extremely high dimensional data which exceed the observation sizes by multiple orders of magnitude. We use cascaded layers of dimensionality reduction techniques combined with regularized regression models for prediction. We present results to predict arrivals to major markets and state wide prices for `Tur' (red gram) crop in Karnataka, India. Our model consistently beats popular ML techniques on many instances. Our model is scalable, time efficient and can be generalized to many other crops and regions. 
We draw multiple insights from the regression parameters, some of which are important aspects to consider when predicting more complex quantities such as prices in the future. We also combine the insights to generate important recommendations for different government organizations.
\end{abstract}

%
%



\maketitle

\section{Introduction}
India is primarily an agriculture-based country and its economy largely depends upon agriculture. According to World Bank, approximately 60 percent of India's land area is used for agricultural purpose making
India the second largest in terms of agricultural land availability. Major chunk of this agricultural land is rain-fed (60\% of total agricultural land) and India ranks first among the rain-fed agricultural countries. Due to unpredictable monsoons there is constant threat to sustainable agricultural production. Presently, contribution of agriculture is about one third of the national GDP and provides employment to over 80\% of Indian population in agriculture and allied activities. Consequently, sustainable development of India largely depends upon the development of agriculture. The agricultural production information is very important for planning and allocation of resources to different sectors of agriculture. 
This is more important for Karnataka, the second largest drought-prone state in India after Rajasthan. The state of Karnataka largely depends on weather conditions for effective farming with more than 75 percent of its arable land in the rainfed (farming by rain water only) regions. Thus, minimizing the impact of natural disaster-related crop losses, particularly from drought, is therefore a major public policy objective for Karnataka government.

The Karnataka Agriculture Price Commission (KAPC) was set up in 2013 to address several agri-related issues.  It works to ensure a) maximum share of consumer price, b) achieve sustainable development in the field of agriculture in the State, c) provide remunerative price for farm produce and d) provide a suitable marketing system
\footnote{http://kapricom.org/downloads/about/termsOfRef-en.pdf}.

Red gram is commonly known as Tur or Arhar (Pigeon pea) in India and is the second most important crop in the pulses category in the country after gram (chana). The ability of red gram to produce high economic yields under soil moisture deficit makes it an important crop in rainfed and dryland agriculture. 
India contributes for nearly 65\% of world's total red gram production. However, it is gaining importance in African countries due to its adaptability to limited moisture conditions and also favorable nitrogen fixation properties exhibited by the crop.


\begin{figure}
    \centering
    \includegraphics[height=4cm,width=4cm]{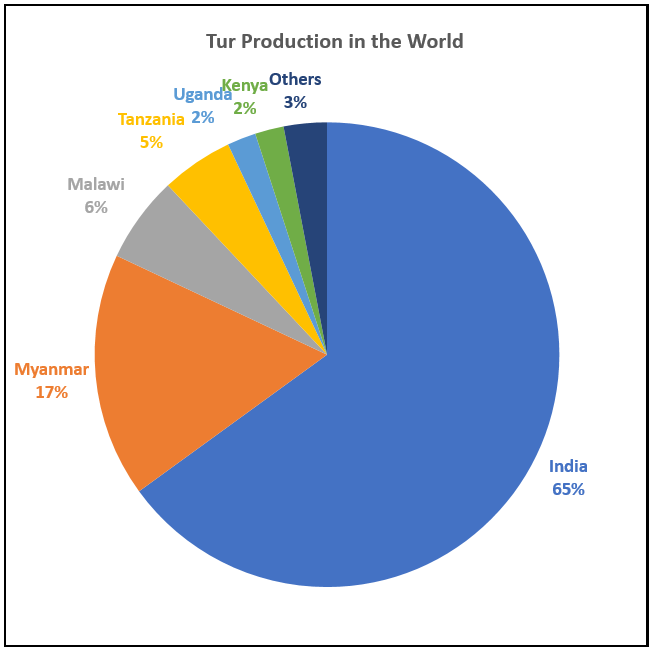}
    \includegraphics[height=4cm,width=4cm]{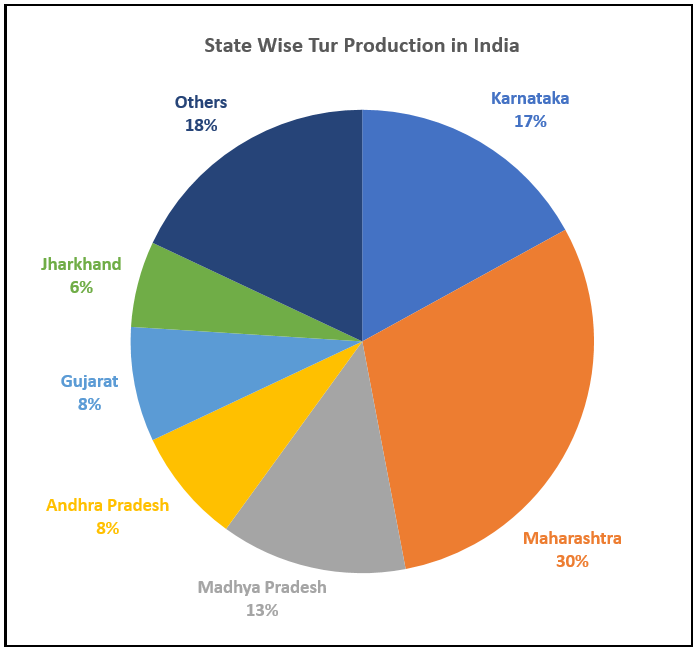}
    \caption{Tur Production data in the world as well as state wise distribution in India.}
    \label{fig:TurDalProp}
\end{figure}

The government of India publishes minimum support prices (MSP) for important agriculture commodities, and Tur is an important crop in that respect. For the last few years Tur has seen huge market disruptions owing to poor prediction of arrivals as well as corresponding MSP. 
There has been sharp variations in prices on a year to year basis and it has led to lot of focus in the prediction of the arrivals to market, both from the government side and also from the agro-research community perspective. 
The government of Karnataka along with KAPC has studied the commodity arrival patterns and the subsequent problem of price prediction problem for several years. 
Arrival prediction helps in identifying the predicted yield for a particular commodity, which further helps in planning the price related interventions. 
KAPC is collaborating with Microsoft to understand and predict these patterns. The model and the subsequent insights described in this work were presented to the state government at the end of last year.\footnote{https://www.indiatoday.in/pti-feed/story/karna-govt-signs-mou-with-microsoft-for-technology-oriented-1074039-2017-10-27\\https://www.techgig.com/tech-news/microsoft-ai-to-help-karnataka-farmers-get-higher-crop-yields-137863\\http://www.thehindubusinessline.com/economy/farmers-look-to-harvest-the-fruits-of-ai/article9928335.ece\\http://indianexpress.com/article/india/karnataka-govt-inks-mou-with-microsoft-to-use-artificial-intelligence-for-digital-agriculture-4909470/}

In this work we describe a 
market wise commodity arrivals prediction (for total arrivals in a month to a market) over several primary markets and also at the entire state level.
We present results against standard methods and show that the remote sensing based method performs better than just market based time-series prediction models or other popular machine learning (ML) based methods.


\subsection{Related Work}
Economists have historically dealt with small sets of data, but this constraint is rapidly changing in the current day and age as new and more detailed data becomes easily available. With the advent of remote sensing, the amount of data that economists can look at went to several gigabytes and ranging over millions of observations. One can now analyze large chunks of data with very little biases / corrections / errors introduced by human manipulations. According to~\cite{Donaldson2016}, the main advantages of remote sensing data to economists can be classified into different categories: 1) access to information which is difficult to obtain by other means; 2) unusually high spatial resolution.

The first advantage arises from the fact that remote sensing satellites can collect panel data at low marginal cost, repeatedly over large time epochs, and usually at very large scale. This data can then be treated as a proxy for a wide range of difficult to measure characteristics. Many topics of potential interest have already been measured remotely and used in other fields such as urban development, building types, road characteristics, pollution indices, beach quality and many more. The ground data needed to study these fields in depth would be prohibitively expensive, and difficult to measure with desirable accuracy. Additionally, there are scenarios when the official government counterparts of some remotely sensed statistics (such as pollution or forestry) may be subject to human manipulation. 

The second advantage of remote sensing data sources is that they are typically available at a substantially higher degree of spatial resolution than are traditional data. Much of the publicly available satellite imagery used by economists provides readings for each of the hundreds of billions of 30m$\times$30m grid cells of land surface on Earth. The freely available MODIS data~\cite{MODIS,MODIS2} 
gives resolutions as 250m, 500m and 1000m. These resolutions are mostly sufficient for econometric studies. Even higher resolution data can be obtained for very targeted applications / studies.


Given the rate at which algorithms for crop classification and yield measurement have improved in recent years, future applications of satellite data are likely to be particularly rich in the agricultural arena. Different bands, and combinations thereof, have various useful properties. 
Plants reflect different sets of frequencies, at different rates, for different stages of their life cycle. For this reason, functions of reflectance in specific portions of the visible and infrared spectra can provide information about vegetation growth. This insight is used to produce the commonly used Normalized Difference Vegetation Index (NDVI)~\cite{Diorio1989}. Infrared data can also be used to measure temperature and, indirectly, precipitation when applied to clouds (for example,~\cite{Novella2013}). Different techniques have been proposed to merge remote sensing data, more specifically NDVI with yield and crop parameter prediction. Time series based methods for yield prediction were proposed by Rembold et al.~\cite{Felix2013}. Supervised classification techniques were proposed by Doraiswamy et al.~\cite{Doraiswamy04}. Linear regression based crop coefficient prediction models were proposed by Kamble et al.~\cite{Kamble2013}. In a recent paper Aviv and Lundsgaard-Nielsen~\cite{Aviv2017} propose a method to predict soy yield by coupling remote sensing data with soil measurement data for 58 locations.

You et al.~\cite{You2017} propose a new method where they go even beyond NDVI and work directly with the band images obtained from satellites to estimate cop yields. Their method is shown to work good for countries like USA due to ready availability of large amounts of clean data. We concentrate solely on data from the state of Karnataka (KA), India, which is available from 2013 onwards on a monthly basis. The amount of data is not enough to train the data intensive neural method described in~\cite{You2017}. Yield estimation is a difficult problem which needs clean data both for modeling as well as validation purposes. You et al.~\cite{You2017} use USDA published data as the ground truth for validating their models. Similar quality data is hard to obtain for developing countries. Moreover, the farm sizes as well as cropping patterns are very different in the developing countries as compared to countries like USA.

In this paper, our contribution is proposed framework to work with limted timeframe data at large scale, a situation often faced in delevoping countries. To our knowledge no work has been done in the mentioned context, especially for any South-Asian countries to predict market arrival and associated insights.

\section{Data}

\subsection{Normalized Difference Vegetation Index (NDVI)}
NDVI is calculated from the differential rates of visible and near-infrared light reflected and absorbed by vegetation. Healthy vegetation absorbs most of the visible light that hits it, and reflects a large portion of the near-infrared light. Unhealthy or sparse vegetation reflects more visible light and less near-infrared light. Nearly all satellite based vegetation indexes employ this difference to quantify the density of plant growth on the Earth. One of the most common indexes, popular with the research community, is the Normalized Difference Vegetation Index (NDVI). NDVI is computed as the ratio between the difference and the summation of the near-infrared (NIR) and the visible radiation (VIS) as shown in Eq.~\ref{Eq:NDVI}. 
\begin{equation}
    \label{Eq:NDVI}
    NDVI = \frac{NIR - VIS}{NIR + VIS}
\end{equation}
where $NIR$ is the reflection for Near Infra Red bands and $VIS$ is the reflection for Visible spectrum. Calculations of NDVI for a given pixel (location on Earth), always results in a number that ranges from minus one (-1) to plus one (+1). However, no green vegetation gives a value close to zero. A zero means no vegetation and close to +1 (0.8 - 0.9) indicates the highest possible density of green vegetation.

In our work, MODIS data was used for the compuatation of NDVI. MODIS uses NIR $(0.7-1.1 \mu m)$ and red $(0.6-0.7 \mu m)$ wavelength bands for the computation of NDVI. 
The spatial resolution of the data is $250m\times 250m$. 
NDVI data is extremely noisy due to the inherent difficulty in capturing the data. 
We smooth the NDVI data by averaging over spatial neighborhoods as shown below.
\begin{equation}\label{Eq:NDVISmooth}
    x_{k,t} = \frac{1}{|N(k)|}\sum_{i\in N(k)}f(z_{i,t})
\end{equation}
where $z_{i,t}$ is the raw NDVI value obtained from satellite images at location $i$ and time $t$, $N(k)$ is the neighborhood of $k$, $x_{k,t}$ is the smoothed transformation output, and $f(.)$ is the transformation function. In the current work we define $f(x) = x$ as the identity function. Putting non-linear combinations for the raw NDVI values is left as a future investigation.

We perform a sampling of the locations to further reduce redundancies in the NDVI data. We divide the geographical region of interest, state of Karnataka, into uniform blocks, and select the centroid of the blocks as the representative locations. These representative locations are selected such that they coincide with the major markets of Tur in Karnataka. Once the locations are identified we perform the transformation and smoothing of the centroid as mentioned in Eq.~\ref{Eq:NDVISmooth}.
\begin{table}
    \centering
\begin{tabular}{l|c|l|c}
    Commodity & Types & Commodity & Types \\ \hline
    Cereals & 15 & Drugs \& Narcotics & 3 \\
    Dry Fruits & 2 & Fibre Crops & 3 \\
    Forest Products & 8 & Fruits & 20 \\
    Live Stock/Poultry & 11 & Oil Seeds & 18 \\
    Other & 12 & Pulses & 21 \\
    Spices & 10 & Vegetables & 47
\end{tabular}
\caption{Commodities and their types tracked by Karnataka government across various markets.}
    \label{fig:CommodityTypes}
\end{table}

\subsection{Market Data}
\begin{figure}
    \centering
    \includegraphics[width=8cm]{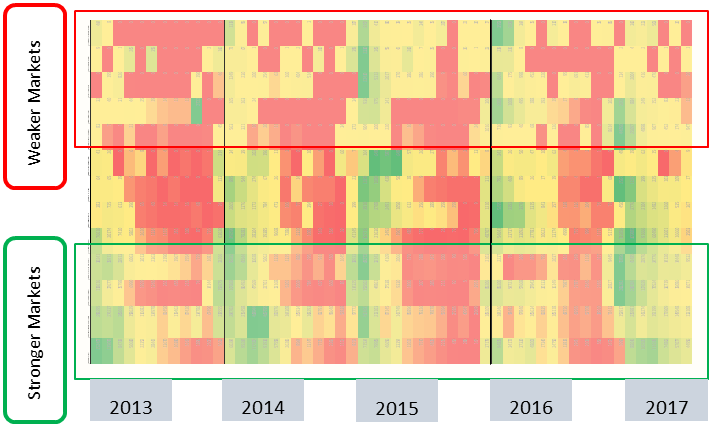}
    \caption{Heatmap for Arrivals to markets over time. The green blocks indicate high monthly arrivals and the red ones indicate low arrivals. Yellow is intermediate. Different rows indicate different markets.}
    \label{fig:marketData}
\end{figure}

Commodity market data for the state of Karnataka is updated daily at the state market portal~\cite{KrishiMVahini}. 
The state government publishes daily arrivals and prices data for the list of commodities as shown in Table.~\ref{fig:CommodityTypes}. The data available from Jan'2013 onward is reliable and relatively noise free.
For each of its 82 markets, the government of Karnataka publishes the commodity variety, grade, arrivals, minimum, maximum and modal prices for each day. Out of the 170 commodities tracked by the state government, the central government of India provides minimum support price (MSP) for only 14 crops with a total of 17 varieties~\cite{MSP}. 
Tur is one of the crops for which the MSP is provided and Karnataka is the second largest producer of Tur in India accounting for around 17\% of total production, which is almost 11\% of the Tur production across the globe. The arrivals distribution over time for some of the 
Tur producing regions of Karnataka are shown in Fig.~\ref{fig:marketData}.

\begin{figure}[ht!]
    \centering
    \includegraphics[width=8cm]{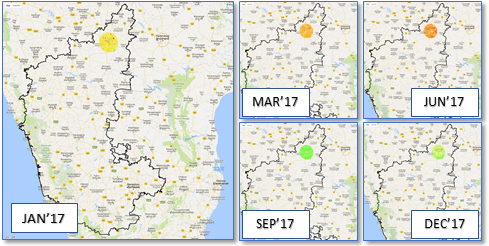}
    \caption{NDVI distribution over time: The state of Karnataka (KA) with the NDVI profile for dominant Tur market Kalaburgi shown in the left pane. The NDVI profile changes distinctly with the season as seen in the remaining panels. Green color dots mean closer to one and greener fields. Yellow is intermediate and Red color dots mean close to zero (no vegetation).}
    \label{fig:ndviDistribution}
\end{figure}

\section{Model}\label{Sec:Model}
Due to historically heavy production of Tur in the northern parts of the state, 
this region is called the TUR bowl of Karnataka. 
The primary thesis of this work is the assumption that 
the NDVI values computed 
for these regions, aggregated over a few years would be a good indicator to predict the arrival of the commodity to the local markets. The change of NDVI over a calendar year for the major crop area of Kalaburgi in Karnataka is shown in Fig.~\ref{fig:ndviDistribution}. The green dots show higher NDVI owing to increased greenery. Yellow dots show intermediate vegetation growth over the same area and red dots show low growth at different times of the year. This pattern is observed for the Kharif season (June / July rainy season sowing).
We have collected NDVI data for more than 6 Million locations distributed all over Karnataka. This data is smoothed according to Eq.~\ref{Eq:NDVISmooth}, where the transformation function $f(x) = x$ is the identity function and the free parameter is the neighborhood size $|N(k)|$. This reduces the number of NDVI data points for the entire state to $6/|N(k)|$ Million locations. We assume that locations situated nearer to the markets provide more useful information than those situated further away. Hence, a set of locations filtered based on their proximity from the markets are selected for further processing. 

Let us assume, a market $M$ at location $c$ and consider all locations $l$ such that the distance $d(l,c)< D$. Our method collects the NDVI values for all the locations $l$ which are close to the market ($D=150-300\mathtt{km}$), and uses them to learn a model to predict the arrival of Tur to the market $M$. The number of distinct locations $L = |l|$ is of the order of $\approx 10-20$K. 
In data cleaning exercise, we used linear inetrpolation to fill few missing observations in market data and removed high variance locatons(more than 75-percentile) among the available set of locations. It was empirically observed that these high variance locations adversely effect future prediction accuracy. After data cleaning, we applied systematic sampling such that for a given market $\approx 7$K locations are available. 

Traditionally, researchers have been using principle component to reduce dimension or ridge regression to avoid non-invertible matrix, which arises in high dimension data or where predictors are likely to have high correlation among each other. In current set-up, we have extremely large dimensions as a set of distinct locations, and near-by locations are likely to have highly correlated NDVI values. We evaluated the performance of these two methods and shown comparative results at the end of the section. We also considered estimating seasonality component but later discarded that thought based on 
arguments provided by agriculture scientists. The main concern was the fact that crop maturity depends heavily on water availability (soil moisture) and appropriate weather conditions, which does not strictly follow seasonal pattern, especially in rainfed regions where significant variations was observed in weather pattern on available last three/four years data. 
Limited historical data (of a few years) along with disrupted weather pattern would adversely effect any estimate of seasonal components and would only add noise in the predictions.

Based on the assumptions mentioned till this point, we explain our proposed regression framework, where first important variables are selected using penalized regression and then principle component regression is applied on selected variables to predict response variables.

Let us assume $\mathbf{x}_t = [x_{1,t}, x_{2,t},\ldots x_{l,t}]'$ is the vector of NDVIs for all locations $l\in L$ falling within a distance $D$ of market $M$ at time $t$. The arrival $y_t$ of Tur for the same market $M$ at time $t$ is then predicted as an elastic-net~\cite{Zou05,Friedman09regularizationpaths} based regression outcome. We collect data for multiple time points into one single optimization and then minimize the following functional:
\begin{eqnarray}
    \label{Eq:ElastiNet}
    \boldsymbol{\mathscr{L}}(\boldsymbol{\beta},\gamma,\lambda;\mathbf{X}_{t-1},\mathbf{Y}_t) &=& \frac{1}{2T} \|\mathbf{Y}_t - \mathbf{X}_{t-1}\boldsymbol{\beta}\|_2^2 \\ \nonumber
    && + ~ \lambda \left[ \frac{1-\gamma}{2} \|\boldsymbol{\beta}\|_2^2 + \gamma \|\boldsymbol{\beta}\|_1 \right]
\end{eqnarray}
where the input matrix $\mathbf{X}_{t-1}\in {R}^{T\times \{L+1\}}$ is the matrix obtained by appending the NDVI data for all the locations near a market for several time instances $T$, and the additional column is all ones to account for the bias factor $\beta_0$. The number of rows $T$ denotes the time window in the past which is used for the regression. The second term in Eq.~\ref{Eq:ElastiNet} is the $L_2$ penalty which penalizes non-smooth combination vector $\boldsymbol{\beta}$ thereby bringing correlated factors together ~\cite{isl01}. The third term with the $L_1$ penalty enforces sparsity in the combination vector, such that only important locations survive in the final outcome prediction. The two constraints together form the elastic-net penalty. The Jacobian for Eq.~\ref{Eq:ElastiNet} with respect to the parameters $\boldsymbol{\beta}$ can now be written as
\begin{equation}
    \label{Eq:JacobianElastiNet}
    \boldsymbol{\mathscr{J}}_{\boldsymbol{\beta}} = - \frac{1}{T}~ \mathbf{X'}_{t-1} \left( \mathbf{Y}_t - \mathbf{X}_{t-1}\boldsymbol{\beta} \right)  +  \lambda (1-\gamma) \boldsymbol{\beta} + \lambda \gamma ~ \mathrm{sign}(\boldsymbol{\beta})
\end{equation}
where $\mathrm{sign}(.)$ gets the element wise sign of the vector argument. The update for the elements of $\boldsymbol{\beta}$ can now be written as shown in~\cite{Friedman07pathwisecoordinate} as
\begin{eqnarray}
{\beta_j} \leftarrow \frac{\mathcal{S} \left( \frac{1}{T}\sum_{i=1}^T x_{ij}( y_i - \hat{y}_i^j ),~ \lambda \gamma \right) }{1+\lambda (1-\gamma)}~~ \forall j \neq 0
\end{eqnarray}
where
\begin{eqnarray}
 \hat{y}_i^j &=& \beta_0 + \sum_{l\neq j}x_{il}\beta_l  \\
 S(z,\gamma) &=& \mathrm{sign}(z).\max( |z|-\gamma,~ 0)
\end{eqnarray}
is the fitted value excluding the contribution from $x_{ij}$, and hence $y_i - \hat{y}_i^j$ is the partial residual for fitting $\beta_j$. $S(z,\gamma)$ is commonly called the soft-thresholding operator. The bias term $\beta_0$ can be obtained by solving the closed form equation
\begin{equation}
-\frac{1}{T}\mathbf{1}' \left( \mathbf{Y}_t - \mathbf{X}_{t-1}\boldsymbol{\beta} \right) - \lambda(1-\gamma) \beta_0 + \lambda \gamma \mathrm{sign}(\beta_0) = 0
\end{equation}
In our formulation more weight is given to $L_2$ penalty $(\gamma <= 0.05)$, to encourage keeping as many important factors as possible. 
High values of $\gamma$ tend to remove too many locations and consequentially the derived features from these sparse locations become unstable. The instability is probably due to fact that when $\gamma$ is large, very few locations are picked to represent a region and therefore derived feature would be based on small sample and hence unstable.  

The output of the elastic-net penalty is several orders lesser in dimension than the initial size of the neighborhood $|l|$. But the number of dimensions to work with is still upto a 1000 or so. We further reduce the dimension by first generating new factors as the principle components (PC) ~\cite{multivariate01} of selected variables and then selecting important PC factors from regression with $L_1$ norm.

Let us assume, without loss of generality, that the first $p$ dimensions of the vector $\boldsymbol{\beta}$ are non-zero and all the rest are  zeros. Selecting the first $p$ corresponding entries from the data vector leads to a modified data vector denoted as
\begin{equation}
    \label{Eq:FirstReduction}
    \hat{\mathbf{x}} = \mathbf{x}^T.\boldmath{\mathrm{1}}_{\beta>0}
\end{equation}
where $\boldsymbol{\mathrm{1}}_{\beta>0}$ denotes the indices where $\boldsymbol{\beta>0}$. 
Next, we perform principle component regression of market arrival series on selected variables with $L_1$ penalty as shown in Eq.~\ref{Eq:ElastiNet4PCA}.
\begin{eqnarray}
    \label{Eq:ElastiNet4PCA}
    \boldsymbol{\mathscr{L}}'(\boldsymbol{\alpha},\lambda;\mathbf{F}_{t-1},\mathbf{Y}_t) &=& \|\mathbf{Y}_t - \mathbf{F}_{t-1}\boldsymbol{\alpha}\|_2^2  + \lambda\|\boldsymbol{\alpha}\|_1 
\end{eqnarray}
where $\mathbf{F}_{t-1}$ is the matrix obtained from stacking $f_i (\hat{\mathbf{x}}_{t-1} )$ which is the projection of $\hat{\mathbf{x}}_{t-1}$ on the $i^{th}$ principal component. Collecting different $y_t$ for different time instances we create a target vector $\mathbf{Y}_t$. 



After this exercise, we are left with very few factors $(|f|\approx 10)$. Once we select important PC factors, we run linear regression with arrivals as dependent variable and the PC factors as predictors as shown in Eq.\ref{Eqn:PCARegression}. Note that $\alpha_i$'s need not be positive as production is postively related to NDVI only certain part of lifecycle of crop. e.g. during harvesting period decline in NDVI followed by large arrival in market. Also, factors associated with water body are negative related, as low NDVI in such regions means good water availability, which is good for crop.

\begin{eqnarray}\label{Eqn:PCARegression}
    {y}_t &=& \alpha_0 + \sum_{i \in |f|}\alpha_i f_{i} (\hat{\mathbf{x}}_{t-1} ) 
\end{eqnarray}

In all above equations, we took $\mathbf{Y}_t$ as log arrival due to the constraint that arrival cannot be less than zero and sensitivity at high and low scale of $\mathbf{Y}_t$ is not same with respect to any changes in factors. We validated these assumptions and ran the experiments for several major Tur markets in Karnataka. 

We compare our proposed model, denoted as RegPCR, against standard tree based methods, RandomForest and Boosting, time series based method ARIMA~\cite{Hyndman08automatictime,timeseries01} and other methods such as ridge and principle component regression. Random forest is known to take care of high dimensional data without over-fitting and with relatively minimum effort in parameter tunning~\cite{randomforest,Statistics01randomforests}. Gradient Boosting (GB)~\cite{Friedman2002,Ridgeway05} is known to perform better than other tree based methods, but tends to over-fit with increasing number of trees. We experimented with GB's parameters (no. of trees, shrinkage and tree depth) and selected best combination. In ARIMA, a univariate arrival time series is modeled and used for prediction for next month. We compute the mean absolute error (MAE) between the predicted arrivals and actual values obtained from the market (Eq.~\ref{Eq:MAE}).
\begin{equation}\label{Eq:MAE}
\mathrm{MAE} ={\frac {\sum _{t=1}^{T}\left|Actual_{t}-Predict_{t}\right|}{T}}
\end{equation}


\begin{figure}[htp!]
    \centering
    \includegraphics[width=8cm]{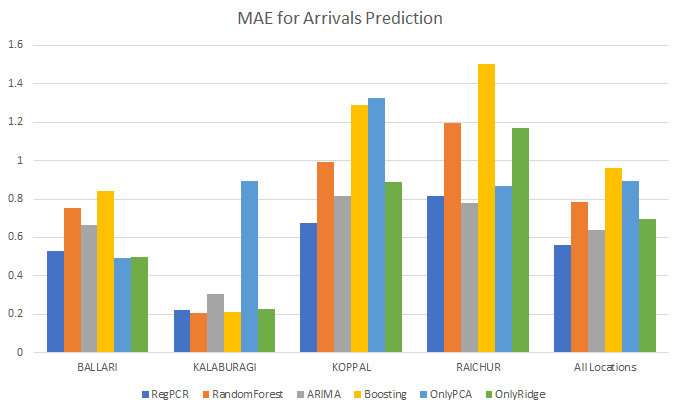}
    \caption{Mean Absolute Error(MAE) for arrival prediction based on our method(RegPCA). We compare against Random Forest, ARIMA, Boosting, PCA and Ridge Regression. The final block shows mean performance across markets.}
    \label{fig:MAP}
\end{figure}
The comparative results are shown in Fig.~\ref{fig:MAP}. Our method performs better than other methods in most markets and is also the clear winner for the aggregated performance. 
Tree based methods - Random forest, Gradient Boosting do not perform well across most markets. It is due to the fact that with such a small sample size computation intensive methods often fail. We also evaluated other parametric methods appropriate in high dimension data like ours. Ridge regression not consistent across markets but perform better than RandomForest and Boosting. We also observed that direct application of Principle component regression perform poorly across markets. This could be due to noise contribution by non-relevant locations.

We explore relation among different market arrivals with total arrival of the commodity in the state. The primary hypothesis is the notion that primary markets within a state are strong indicators of the total arrival to the state. 
Consequently, we use the predicted arrivals for the individual markets at Ballary, Kalaburagi, Koppal, and Raichur to predict the total arrival of Tur for Karnataka. 
\begin{equation}
    A_{KA} = \alpha_0 + \sum_i \alpha_i A_i
\end{equation}
where $\{i \in \mathtt{[Ballary, Kalaburagi, Koppal, Raichur]}\}$. The simple regression model for total arrival in Karnataka $A_{KA}$, gives an adjusted-R-square $> 98\%$ with a p-value $< 0.001$. 


\begin{figure*}[htb]
    \centering
    \includegraphics[width=15cm]{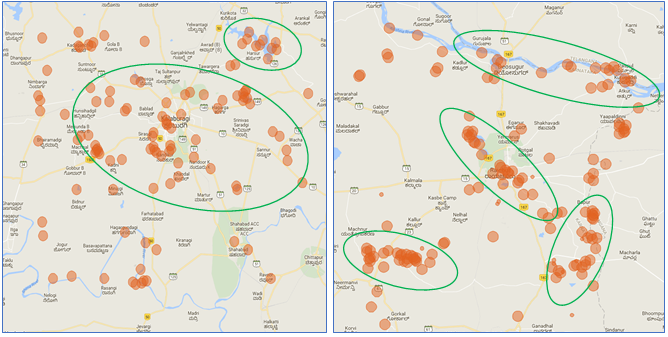}
    \caption{Important Regions identified by our method for Kalaburgi (left) and Raichur (right) market. Note that for the stronger market Kalaburgi in the left panel, lots of important locations are scattered in the region of interest. For relatively weaker markets such as Raichur, we find lesser scattering of locations. Each circle is a location, Size represents importance as predicted by our method. Color is for ease of visualization.}
    \label{fig:RegionAnalysis}
\end{figure*}

\section{Data driven Insights from Market Analysis}
\subsection{Location Identification}
We analyzed the models presented in Sec.~\ref{Sec:Model}, to derive useful insights from the commodity arrival prediction work-flow. For each market, we selected important variables (latitude-longitude pairs) identified by our model. Since we predict for multiple times, we accumulate the importance of a region based on how important it was for one particular prediction (regression coefficient magnitude) and also weight it with the number of times this location was dominant across the various predictions. 
We selected important variables which are consistently picked by elastic-net regression and have large absolute regression coefficients (Eq.~\ref{Eq:ElastiNet}, Eq.~\ref{Eq:FirstReduction}). 

We take two markets to explain our findings. Kalaburgi, the strongest Tur market and Raichur - relatively weak market which tends to grow more than one crop in same season. For Kalaburgi, the important locations are distributed over entire region, thereby pointing to the fact that all the locations are similar in their importance. 
On the other hand Raichur provide an interesting picture where we find multiple clusters - some are around water bodies and other are large arable land. Note that in Raichur important locations are restricted to only some part of region, highlighting locations which are contributing in Tur production. It also implies other arable locations which are not selected as important contribute in different agri-commodity production, not in Tur. The locations with their relative importance (circle size and color for disambiguation) are shown in Fig.~\ref{fig:RegionAnalysis}. 

Our method predicts strong region association with the locations which are close to water bodies. This pattern is replicated across the other locations as well, but fades when the market is not among top producer of the crop itself. 
More explorations along these lines has been marked as a future research activity. 


\subsection{Crop-Cutting Experiments}
One of the important problems which the government in India is facing is to come up with better locations for the Crop-Cutting Experiments (CCEs)~\cite{CCE}.
The current guidelines for performing CCEs have the following salient features
\begin{itemize}
    \item Crop Cutting Experiments (CCE) under scientifically designed General Crop Estimation Surveys (GCES).
    \item Around 950 thousand CCEs across India.
    \item Stratified multistage random sampling: Tehsil / Taluk $>$ Revenue Village $>$ Survey Number / Field $>$ Experimental Plot (Specified size / shape)
    \item 80$-$120 CCEs for a crop in a major district.
\end{itemize}
Although the strategy to perform the CCEs looks well motivated from the salient features, the survey results done later tell a different story~\cite{KAConf}. 
According to the published work, the CCE estimates for Tur, performed by the central government agency and the state government agencies differed from $+11.89\%$ to $-126.13\%$! Many of these discrepancies have been attributed to the sampling system which can be also influenced by external factors. Our method can identify the important locations for CCEs for the next cycle based on historical evidence and remote sensing data, thereby ensuring transparency as well as equitable gains. Additionally, identified farm lands in important locations can also be used as investigating points to identify real reason (soil quality, choice of fertilizer, farm management practices) for their consistent performance and replicating it as best practice across regions.

\subsection{Price Prediction}
Agriculture commodities arrive to the market and are sold to buyers on various days in a month. For each day, the government records the mode, min and max price for each commodity over each market. There are some differences in price distribution from market to market, and the individual daily distributions across the markets are noisy as well. To make the price predictions useful for the state government's policy making efforts, we propose to predict the monthly average price for the state instead of predicting daily variations for each market. We use the predicted arrivals to model the commodity price in the market. 

\begin{figure}
    \centering
    \includegraphics[width=8cm]{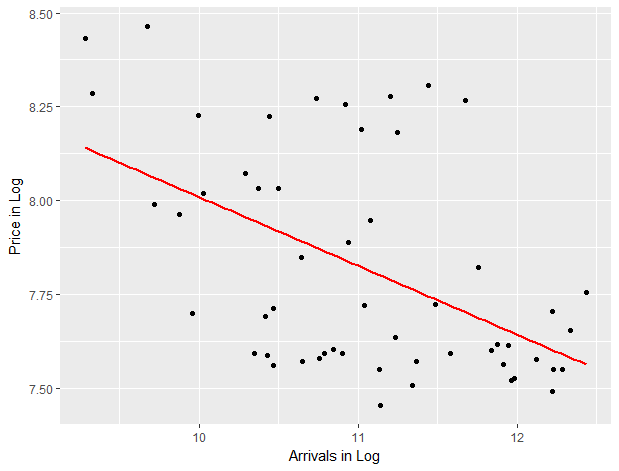}
    \caption{Price vs Arrivals for Tur in Karnataka from Jan'13-Nov'17.} 
    \label{fig:PrvsRawArr}
\end{figure}

It is a well known principle of macroeconomics that as the commodity arrival increases / decreases price goes down / up respectively.
Generally, a convex function describes the relationship between price (P) and supply/arrivals (A), given by 
\begin{eqnarray}\label{Eq:marketRel}
    && P^\alpha A^\beta = \mathcal{C} 
\end{eqnarray}
where $\mathcal{C}$ is a non-negative constant. 
In Fig.~\ref{fig:PrvsRawArr}, each observation represents monthly arrival and the corresponding price for Tur, for the state of Karnakata. The simple relationship in Eq.~\ref{Eq:marketRel} does not hold for extended periods of time. It has been observed under various contexts that due to seasonality and many other factors the curve can shift or transform in various ways.  

A closer look at Fig.~\ref{fig:PrvsRawArr}, shows that there is negative correlation between price and arrival, though price conditioned on arrival has high variance. This is due to the fact that major part of commodity which arrives during the harvest season, is consumed all over the year, and therefore, high surge in arrivals may not reflect proportional drop in price and vice-versa. The price is also affected by irregular interventions by the government either by declaring changes in minimum support prices to support farmers in case of over supply or by announcing import initiative or export ban to cool off inflation due to crop failure.

Let $\{p_t\}_{t=1}^T$ be price time series and $\{a_t\}_{t=1}^T$ is log arrival time series. To incorporate consumption effect we take exponential weighted average on past observations.
We define modified arrival at time t as
\begin{equation}
    A_t = f_w(a_t)
\end{equation}
\begin{equation}
    P_t = log(p_t)
\end{equation}
where 
\begin{equation}
    f_w(a_t) = \sum_{i=1}^{12}a_{t-i+1} w^{i-1}
\end{equation}
such that $w$ is the exponential decay factor and the index $i$ runs over last twelve months.
Next, we take differences with duration $d$, and define
\begin{eqnarray}
    A_t^d &=& A_t - A_{t-d} 
\end{eqnarray}
\begin{figure}
    \centering
    \includegraphics[width=8.5cm]{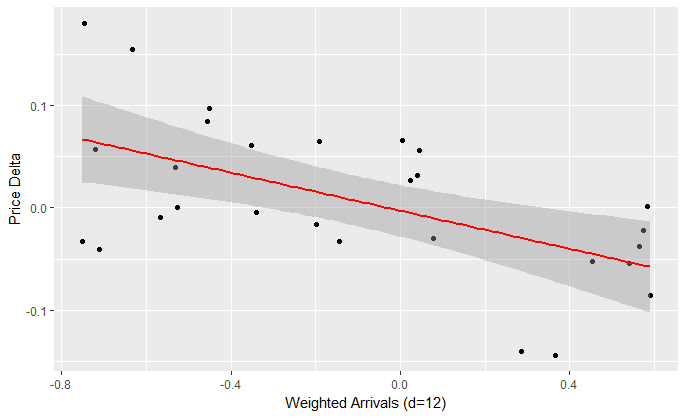}
    \caption{Price Difference ($\Delta P_t$) vs Weighted Arrivals Difference ($A_t^{12}$) on available data. }
    \label{fig:PricePrediction}
\end{figure}
For $d=12$, the above equation calculates extra commodity arrived at month t compared to last year. This additional commodity should have effect on the price variations observed for the current and coming months. We model $\Delta P_t$ = $P_t$ - $P_{t-1}$ as a function of $A_t^d$, where d=12. The variation of $\Delta P_t$ over the weighted arrivals is shown in Fig.~\ref{fig:PricePrediction}. 

\begin{figure}
    \centering
    \includegraphics[width=8cm,height=6cm]{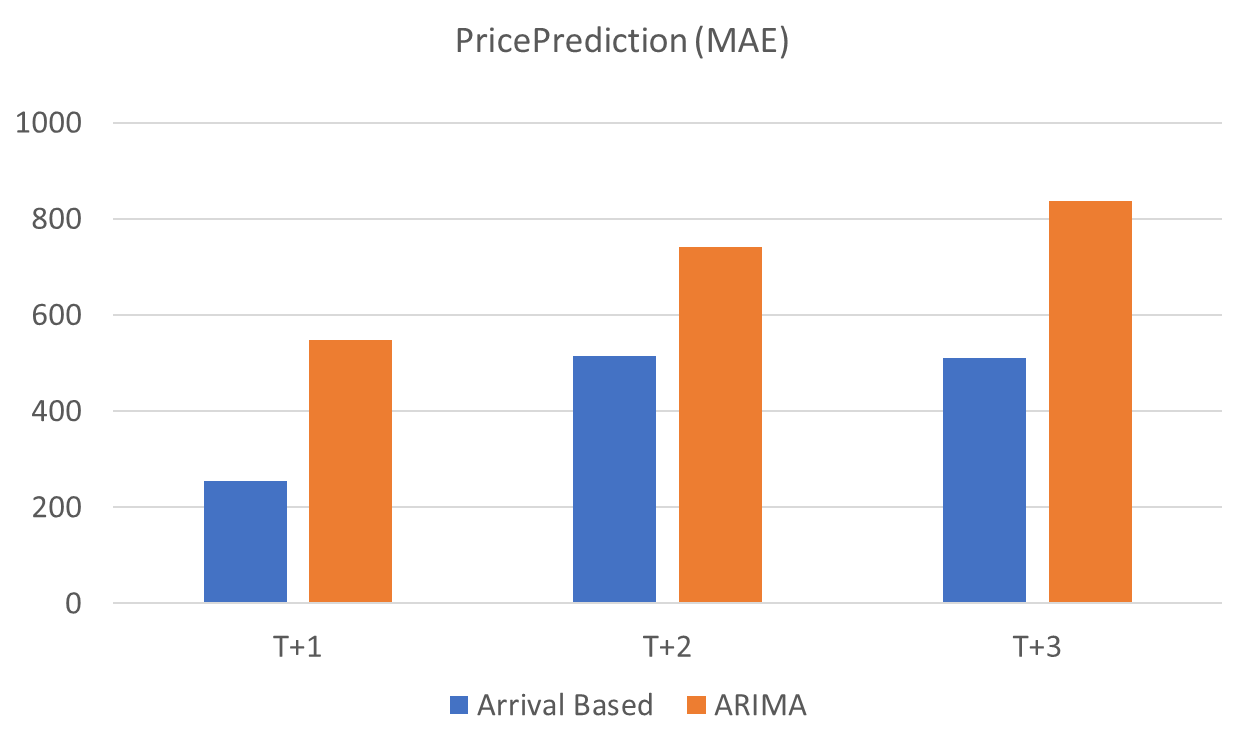}
    \caption{Price prediction comparison. Arrivals based method proposed in this work against a standard ARIMA based model.}
    \label{fig:pricePred}
\end{figure}

Once we predict arrival for the next month $\hat{A}_{t+1}^{12}$, we can predict the corresponding price for the next month and for a few more months in the future. 
Let us assume we want to predict the difference in price, $\Delta P_{t+1}$, based on realized arrivals $A_t^{12}$ available till time $t$, and the estimated arrival $\hat{A}_{t+1}^{12}$. 
Eq.~\ref{Eq.PricePred} denotes this prediction task for the next three months indexed by $k\in [1,2,3]$.
\begin{eqnarray}\label{Eq.PricePred}
    \Delta P_{t+k}  =  a_0^k + \overbrace{ a_1^k \hat{A}_{t+1}^{12}}^\text{estimated factors} + \overbrace{a_2^kA_{t}^{12} + a_3^kA_{t-1}^{12}}^\text{realized factors}
\end{eqnarray}
We observed that for near term predictions our method works well, possibly because it captures monthly price movement and also takes care of annual commodity availability. 
The comparison of our model against ARIMA model over last 12 months is shown in Fig.~\ref{fig:pricePred}. As we move from predicting the prices for the next month, to upto 3 months in the future, the errors by both the models increase. 


\section{Conclusion}
In this paper, we propose a regression framework to predict arrival quantity of Tur to the major markets of Karnataka, India. The proposed method works for extremely unbalanced data setup, where there are less than a hundred observations, but the dimension of those observations can range in millions. 

Our entire approach was motivated by the fact that developing nations do not have large amounts of clean recorded data, but these can still be used in principled frameworks to derive value out of them. We use the extremely limited market data and infuse it with large amounts of remote sensing data to arrive at the unbalanced system of equations which is then solved to predict market factors. We utilize the predicted commodity arrivals to draw important insights which can potentially be used by the policy makers to design better optimized crop-cutting experiments and modify farm practices by identifying important attributes for consistent performance. We were able to use the predicted arrivals to forecast the corresponding price outlook for three months in the future. Our results show that we outperform traditional methods like Random Forest, Boosting, Ridge Regression, ARIMA and Principle component regression by considerable margins. We ruled out the application of neural techniques due to the extreme scarcity of data. We propose to investigate price prediction in more detail as it is an important problem by itself. We also propose to continue studying the key factors found in final model and explore HSI based analysis. 

\section{Acknowldegment}
We would like to acknowldege Niranjan Nayak, Prashant Gupta, Sushil Chordia and KAPC chairman T.N. Prakash Kammardi for their guidance and support for this project.

\bibliographystyle{IEEEtran}
\bibliography{reference}

\end{document}